\documentstyle[12pt,epsfig]{article}
\def\lsim{\raise0.3ex\hbox{$<$\kern-0.75em\raise-1.1ex\hbox{$\sim$}}}
\def\gsim{\raise0.3ex\hbox{$>$\kern-0.75em\raise-1.1ex\hbox{$\sim$}}}

\def\beq{\begin{equation}}
\def\eeq{\end{equation}}
\def\bea{\begin{eqnarray}}
\def\eea{\end{eqnarray}}
\def\bq{\begin{quote}}
\def\eq{\end{quote}}

\def\gappeq{\mathrel{\rlap {\raise.5ex\hbox{$>$}}
{\lower.5ex\hbox{$\sim$}}}}

\def\lappeq{\mathrel{\rlap{\raise.5ex\hbox{$<$}}
{\lower.5ex\hbox{$\sim$}}}}

\def\Toprel#1\over#2{\mathrel{\mathop{#2}\limits^{#1}}}

\begin{document}
\pagestyle{empty}
\begin{center}
  {\bf THE QCD POMERON IN ULTRAPERIPHERAL HEAVY ION COLLISIONS: V. DOUBLE VECTOR MESON PRODUCTION IN THE BFKL APPROACH}
\\

\vspace*{1cm}
 V.P. Gon\c{c}alves $^{1}$, M.V.T. Machado  $^{2,\,3}$, W. K. Sauter $^{3}$ \\
\vspace{0.3cm}
{$^{1}$ High and Medium Energy Group (GAME), \\
Instituto de F\'{\i}sica e Matem\'atica,  Universidade
Federal de Pelotas\\
Caixa Postal 354, CEP 96010-090, Pelotas, RS, Brazil\\
$^{2}$ \rm Universidade Estadual do Rio Grande do Sul - UERGS\\
 Unidade de Bento Gon\c{c}alves. CEP 95700-000. Bento Gon\c{c}alves, RS, Brazil\\
$^{3}$ \rm High Energy Physics Phenomenology Group, GFPAE  IF-UFRGS \\
Caixa Postal 15051, CEP 91501-970, Porto Alegre, RS, Brazil}\\
\vspace*{1cm}
{\bf ABSTRACT}
\end{center}

\vspace*{1.5cm} \noindent

\vspace*{1.3cm} \noindent \rule[.1in]{17cm}{.002in}

\vspace{-3.5cm} \setcounter{page}{1} \pagestyle{plain} 

In this work the double vector meson  production in ultraperipheral heavy ion collisions is investigated within the BFKL approach. The integrated cross sections and event rates  for the processes $AA \rightarrow V_1 V_2 \,AA$ ($V_i = \rho, \omega, \phi, J/\Psi, \Upsilon$)  are computed and theoretical estimates for scattering on both  light and heavy nuclei are given for energies of RHIC and LHC.   \vspace{0.5cm}

\vspace{1cm}

\section{Introduction}
 Photon collisions offer unique possibility to probe QCD in its high energy limit \cite{kw}. The simplicity of the initial state and the possibility of studing  several different combinations of final states make this process very useful for studying the QCD dynamics in  the limit of high center-of-mass energy $\sqrt{s}$  and fixed momentum transfer $t$. In principle, in this regime the QCD dynamics is determined by the QCD Pomeron, with the evolution described by the BFKL equation \cite{bfkl}.  One typical process where the QCD Pomeron can be tested is  the vector meson pair production in $\gamma \gamma$ collisions \cite{ginzburg}.  For heavy vector mesons, this cross section can be calculated using the perturbative QCD methods. First calculations considering the Born two-gluon approximation have been done in Refs. \cite{ginzburg} and after considering the solution of the BFKL equation \cite{motyka}. Recently, we have studied the double vector meson production in $\gamma \gamma$ collisions, where the the color singlet $t$-channel exchange carries large momentum transfer \cite{vic_sauter}.  The presence of a large momentum transfer allow us to calculate the light and heavy vector mesons production cross sections for real photon interactions. We have considered the non-forward  solution of the BFKL equation  at high energy and large momentum transfer \cite{Lipatov} and  estimated the total and differential cross section for the process $\gamma \gamma \rightarrow V_1 V_2$, where $V_1$ and $V_2$ can be any two vector mesons  ($V_i = \rho, \omega, \phi, J/\Psi, \Upsilon$).  Our results indicate that this process could be used in the future $\gamma \gamma$ colliders to constrain the QCD dynamics.

\begin{figure}[t]
\centerline{\epsfig{file=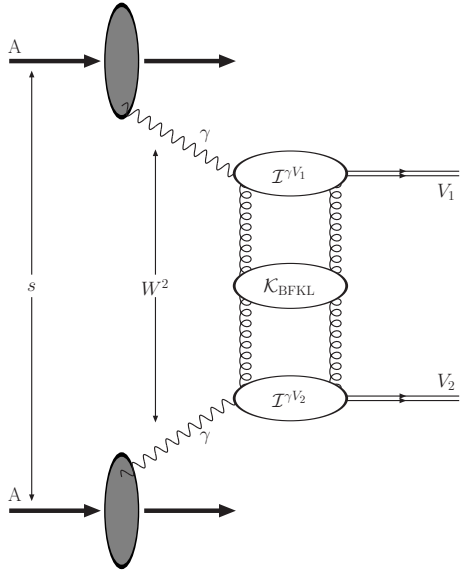,width=6cm}}
\caption{\it The process $AA\rightarrow V_1V_2AA$ and the corresponding kinematical variables. The two-photon scattering is sketched, showing the photon-meson impact factors ${\cal{I}}^{\,\gamma V_{1,2}}$ and the  BFKL Green's function ${\cal{K}}_{\mathrm{BFKL}}$. The grey blobs represent the colliding nuclei.}
\label{fig1}
\end{figure}

On the other hand,  in  Refs. \cite{per1,per2,per3,per4} we have shown that several processes which can be observed in ultraperipheral heavy ion collisions (UPC's) at RHIC and LHC  can be used to constrain the QCD dynamics. In these collisions  the ions do
not interact directly with each other and move essentially
undisturbed along the beam direction. The only possible
interaction is due to the long range electromagnetic interaction
and diffractive processes (For a review see, e. g. Refs.
\cite{bert,bert2,review_mesons}). 
Consequently, it is possible to study photon-nucleus and/or two-photon interactions in these coherent collisions. 
An important aspect which should be emphasized  is that 
relativistic heavy-ion collisions are a potentially prolific
source of $\gamma \gamma$ collisions at high energy colliders. The
advantage of using heavy ions is that the cross sections varies as
$Z^4 \alpha^4$ rather just as $\alpha^4$. Moreover, the maximum
$\gamma \gamma$ collision energy  $W_{\gamma \gamma}$ is $2\gamma
/R_A$,  about 6 GeV at RHIC and 200 GeV at LHC, where $R_A$ is the
nuclear radius and $\gamma$ is the center-of-mass system Lorentz
factor of each ion. In particular, the LHC will have a significant
energy and luminosity reach beyond LEP2, and could be a bridge to
$\gamma \gamma$ collisions at a future $e^+ e^-$ linear collider.
For two-photon collisions, the cross section for the reaction $AA
\rightarrow AA \,V_1 \, V_2$ will be given by (See Fig. \ref{fig1})
\begin{eqnarray}
\sigma_{AA \rightarrow AA \,V_1 \, V_2} = \int \frac{d
\omega_1}{\omega_1} \, n_1(\omega_1) \int \frac{d
\omega_2}{\omega_2}\, n_2(\omega_2) \,\sigma_{\gamma \gamma
\rightarrow V_1 \, V_2} (W = \sqrt{4 \omega_1 \omega_2}\,)
\,\,,
\end{eqnarray}
where the photon energy distribution $n(\omega)$ is calculated
within the Weizs\"acker-Williams 
approximation \cite{bert}. In general, the  total cross section $
AA \rightarrow AA \,\gamma \gamma \rightarrow AA\, X$, where $X$ is
the system produced within the rapidity gap, factorizes into the
photon-photon luminosity $d{\cal{L}}_{\gamma
\gamma}/d\tau$ and the cross section of the $\gamma \gamma$
interaction,
\begin{eqnarray}
\sigma_{AA \rightarrow AA \,V_1 \, V_2}(s) = \int d\tau \, \frac{d
{\cal{L}}_{\gamma \gamma}}{d\tau} \, \hat \sigma_{\gamma \gamma
\rightarrow V_1 \, V_2}(\hat s), \label{sigfoton}
\end{eqnarray}
where $\tau = {\hat s}/s$, $\hat s = W^2$ is the square of the
center of mass (c.m.s.) system energy of the two photons and $s$
of the ion-ion system. The $\gamma \gamma$ luminosity, $d\, {\cal{L}}_{\gamma \gamma}\,(\tau)/d\tau$ , is given by
the convolution of the photon fluxes from two ultrarelativistic
nuclei. Here, we consider
the photon distribution of Ref. \cite{cahn}, providing a photon
distribution which is not factorizable. The authors of \cite{cahn}
produced practical parametric expressions  for the differential
luminosity by adjusting the theoretical results. The comparison with the complete form is consistent within a few percents. The approach given above excludes possible final state interactions of the produced particles with the colliding
particles, allowing reliable calculations of UPC's.

In order to estimate the 
double meson production  in UPC's  it is only necessary to consider a suitable  QCD model for the subprocess at photon level. In Refs. \cite{per1,double_meson} we have analyzed several distinct scenarios for the double $J/\Psi$, $\rho J/\Psi$ and double $\rho$ production, respectively. In particular, in Ref. \cite{per1} we have estimated the double $J/\Psi$ production considering the solution of the BFKL equation at zero momentum transfer and using  a small $t$ approximation in order to calculate the total cross section at  photon level. Moreover, we have evaluated the impact of non-leading corrections to the BFKL equation and compared our predictions with those obtained at Born two-gluon approximation. On the other hand, in Ref. \cite{double_meson} we have estimated the  $\rho J/\Psi$ production considering the approach proposed in Ref. \cite{motyka_ziaja}, where the $\gamma \gamma \rightarrow  \rho J/\Psi$ cross section is given in terms of the gluon distribution on the light meson. Moreover, the double $\rho \rho$ production has been estimated using the pomeron-exchange factorization theorem \cite{Gribov:ga}, which allows relate the  
$\gamma \gamma \rightarrow  \rho \rho$ cross section with those for $J/\Psi J/\Psi$ and $\rho J/\Psi$ production. 
Our main result in those analyzes was to show that the QCD dynamics implies a large enhancement of the cross sections, which become these processes observable in future $AA$ colliders  and the ultraperipheral heavy ion collisions an important alternative to study the QCD dynamics at high energies. It strongly motivates a more detailed study of the double meson production in UPC's considering  more precise theoretical  formalisms for the $\gamma \gamma \rightarrow V_1 V_2$ process, as well as to obtain predictions for other combinations of vector mesons in the final state.

Here, we consider the approach proposed in Ref. \cite{vic_sauter} where the double vector meson production has been estimated in terms of the non-forward solution of the BFKL equation \cite{Lipatov}. It is based on the approach introduced in Refs. \cite{FR,BFLW} to describe the process $\gamma p \rightarrow V X$ at large $t$, which reproduce the measured differential cross sections for longitudinal $\rho$ meson and for $J/\Psi$ production very well \cite{FP,jhep}.  In what follows (Section \ref{sec2}),  we shortly present the main formulae of the approach proposed in Ref. \cite{vic_sauter} and its further application for the double meson production at UPC's. We analyze the theoretical estimates for the energies of RHIC and LHC and its dependence on atomic nuclei. The event rates are also computed, giving the magnitude of the contributions to the possible final states at the coherent collisions. Finally, in Section \ref{conc} we discuss the main backgrounds for the photon-photon process in UPC's and summarize our  results. 

\section{Vector meson pair production in the BFKL approach}
\label{sec2}

\begin{table}[t]
\begin{center}
\footnotesize
\begin{tabular}{||c|c|c|c||}
\hline
\hline
COLLIDER & ION        & $\sqrt{S_{NN}}$ (GeV) & ${\cal L}_{AA}$ (cm$^{-2}$s$^{-1}$)  \\ \hline
\hline
{\bf RHIC} & Si      & 250 & $4.4\times10^{28}$   \\
& Au      & 200 & $2.0\times10^{26}$     \\
\hline
\hline
{\bf LHC} & Ca      & 7200 & $5.2\times 10^{29}$  \\
& Pb      & 5500 & $4.2\times10^{26}$  \\
\hline
\hline
\end{tabular}
\end{center}
\caption{\it Luminosities and beam energies for $AA$  collisions
at RHIC and LHC.}
\label{lum}
\end{table}

In the last few years several studies has analyzed  the double meson production in two-photon processes  considering different approximations for the QCD dynamics \cite{motyka,motyka_ziaja,per1,vic_sauter,Enberg,DIvanov}. In particular, in our previous paper in Ref. \cite{per1}, we have performed a phenomenological analysis for the double $J/\Psi$ production using the forward LLA BFKL solution. In that case, the hard scale was set by the charm quark mass. There, we also studied the possible effects of corrections at next to leading approximation (NLA) level to the BFKL kernel investigating the slow down of the effective hard Pomeron intercept. Afterwards, in Ref. \cite{vic_sauter} the non-forward solution was considered for a larger set of possible vector meson pairs, where the large $t$ values provide the perturbative scale. In that paper the double vector meson production in real photon interactions was studied, the $t$-dependence of the differential cross section was analyzed in detail and the total cross section for different combinations of vector mesons was calculated using the leading order impact factors and  BFKL amplitude. More recently, two other  studies on the process $\gamma^*\gamma^*\rightarrow VV$ has appeared in literature \cite{Enberg,DIvanov}. In the first one \cite{Enberg}, the leading order BFKL amplitude for the exclusive diffractive two-$\rho$ production in the forward direction is computed and the NLA corrections are estimated using a specific resummation of higher order effects.  In the last paper \cite{DIvanov}, the amplitude for the forward electroproduction of two light vector mesons in NLA is computed. In particular, the NLA amplitude is constructed by the convolution of the $\gamma^*\rightarrow V$ impact factor and the BFKL Green's function in the $\overline{\mathrm{MS}}$ scheme. In addition, a procedure to get results independent from the energy and renormalization scales has been investigated within NLA approximation. In what follows we shall consider the approach proposed in Ref. \cite{vic_sauter}, which is valid for real photon interactions and has been compared with the HERA data when applied for photon-proton collisions. As it is based on the leading logarithmic approximation (LLA) for the impact factors and BFKL amplitude, we should consider our results as an upper bound for the total cross sections.

Following Ref. \cite{vic_sauter} we have that the differential cross section for the process $\gamma \gamma \rightarrow V_1 V_2$ is given by 
\begin{equation}
\frac{d\sigma (\gamma \gamma \rightarrow V_1 V_2)}{dt} \; = \;
\frac{16\pi}{81 t^4}
|{\mathcal{F}}_{\mathrm{BFKL}}(z,\tau)|^{2}\,\,,
\label{dsdtgq}
\end{equation}
where $z = (3\alpha_{s}/2\pi) \ln( W^2/\Lambda^{2})$,  $\tau = |t|/M_{V}^{2}$, $M_{V}$ is the mass of the vector meson and $\Lambda^{2}$ is a characteristic  scale related to $M_V^2$  and $|t|$ (See Ref. \cite{vic_sauter} for details). Moreover, in the leading logarithm approximation  (LLA) the coupling constant is fixed and  we assume $\alpha_s$ = 0.2 in our calculations, which is determined from a fit to the HERA data (for details see \cite{jhep}). 
The leading order BFKL amplitude  and  lowest conformal spin ($n=0$), is given by~\cite{Lipatov}
\begin{equation}
\label{BFKLa}
{\mathcal{F}}_{\mathrm{BFKL}}(z,\tau)=\frac{t^{2}}{(2 \pi)^{3}}\int d\nu \frac{\nu ^{2}}{(\nu ^{2}+1/4)^{2}}e^{\chi (\nu )z}I_{\nu }^{\gamma V_1}(Q_{\perp })I^{\gamma V_2}_{\nu }(Q_{\perp })^{\ast },
\end{equation}
 where $Q_{\perp}$ is the momentum transfered, $t=-Q_{\perp}^2$, (the subscript denotes two dimensional transverse vectors) and 
\begin{equation}
\chi (\nu )=4{\mathcal{R}}\mathrm{e}\biggl (\psi (1)-\psi \bigg (\frac{1}{2}+i\nu \bigg )\biggr )
\end{equation}
is proportional to the BKFL kernel eigenvalues~\cite{Jeff-book}, with $\psi(x)$ being the digamma function. 
\begin{table}[t]
\begin{center}
\footnotesize
\begin{tabular}{||c||l|l||l|l||}
\hline \hline
{\bf $V_1V_2$} & {\bf RHIC} & & {\bf LHC} &  \\ \hline 
 & {\bf Si} & {\bf Au} & {\bf Ca} & {\bf Pb} \\\hline
$\rho\rho$ &  $10^{-2}$ (0.42) & 1.2 (0.2) & 4.3 ($2 \times 10^{3}$) & 432 (181)\\
\hline
$\omega\omega$ &  $6 \times 10^{-5}$ ($3 \times 10^{-3}$) & $8 \times 10^{-3}$ ($10^{-3}$) & 0.03 (15) & 3 (1.2)\\
\hline
$\phi \phi$ & $2 \times 10^{-3}$ (0.01) & 0.02 ($4 \times 10^{-3}$) & 0.14 (74) & 14 (6)\\
 \hline
$\rho \omega$ & $7 \times 10^{-4}$ (0.03) & 0.1 (0.02) & 0.3 (174) & 34 (14) \\
\hline
$\rho \phi$ & $10^{-3}$ (0.06) &  0.16 (0.03) & 0.7 (378) & 73 (31)  \\
\hline
$\omega\phi$ & $6 \times 10^{-4}$ (0.03) &  0.013 ($3 \times 10^{-3}$) & 0.06 (32) & 6 (3)\\
\hline
 \hline \hline
\end{tabular}
\end{center}
\caption{\it Double light mesons: The total cross section (number of events/month) for double vector meson production in two-photon reactions in UPC's at RHIC and LHC energies ($t_{min} = 1$ GeV$^2$). Cross sections in units of nb. }
\label{tab1}
\end{table}

 The quantities $I_{\nu }^{\gamma V_i}$ are given in terms of the impact factors ${\mathcal{I}}_{\gamma V_i}$ and the BFKL eigenfunctions as follows \cite{BFLW},
\begin{eqnarray}
I_{\nu}^{\gamma V_i}(Q_{\perp }) & = & -{{\mathcal{C}}_i}\, \alpha_s \frac{16\pi}{Q_{\perp }^{3}}\frac{\Gamma (1/2-i\nu )}{\Gamma (1/2+i\nu )}\biggl (\frac{Q_{\perp }^{2}}{4}\biggr )^{i\nu }\int _{1/2-i\infty }^{1/2+i\infty }\frac{du}{2\pi i}\biggl (\frac{Q_{\perp }^2}{4 M_{V_i}^2}\biggr )^{1/2+u}\\
 &  & \times\frac{\Gamma ^{2}(1/2+u)\Gamma (1/2-u/2-i\nu/2)\Gamma (1/2-u/2+i\nu/2)}{\Gamma (1/2+u/2-i\nu /2)\Gamma (1/2+u/2+i\nu /2)},\nonumber \label{IV} 
\end{eqnarray}
where ${{\mathcal{C}}_i} = {3\Gamma_{ee}^{V_i}M_{V_i}^{3}}/{\alpha_{\mathrm{em}}}$. Notice the non-relativistic approximation for the impact factor is considered here.  The differential cross section can be directly calculated substituting the above expression in Eq. (\ref{BFKLa}) and evaluating numerically the integrals.
The total cross section will be given by
\begin{eqnarray}
\sigma (\gamma \gamma \rightarrow V_1 V_2) = \int_{|t|_{min}}^{\infty} d|t| \,\,\, \frac{d\sigma (\gamma \gamma \rightarrow V_1 V_2)}{d|t|} \,\,,
\label{sigtot}
\end{eqnarray}
where $|t|_{min}$ is the minimum momentum transfer in each particular process.
In order to use  a perturbative approach for the interaction of real photons the presence of a hard scale is necessary. It can be the mass of the vector meson or the momentum transfer in the process. Consequently, for heavy vector meson production we can assume $|t|_{min} = 0$, while for light vector meson production we need to consider a lower limit for the momentum transfer. Here, we will assume   $|t|_{min} = 1$ GeV$^2$ as in Ref. \cite{vic_sauter}. In principle, for smaller values of $t$ non-perturbative contributions should dominate the total cross sections. In what follows we estimate the double vector meson production assuming $|t|_{min} = 0$ when a heavy vector meson is present and $|t|_{min} = 1$ GeV$^2$ for double light vector meson production. For comparison, we also present our predictions for the latter process assuming  $|t|_{min} = 0$, which must be used with some caution, since we are extrapolating the perturbative approach for very small momentum transfer where its direct application is questionable. However, as we expect that the $t$-dependence will not be strongly modified at small $t$, it can be used as a rough estimate.
\begin{table}[t]
\begin{center}
\footnotesize
\begin{tabular}{||c||l|l||l|l||}
\hline \hline
{\bf $V_1V_2$} & {\bf RHIC} & & {\bf LHC} &  \\ \hline 
 & {\bf Si} & {\bf Au} & {\bf Ca} & {\bf Pb} \\\hline
$\rho\rho$ & $3.2$ (140) & 340 (68) & 2821 ($10^6$) & $2 \times 10^5$ ($10^5$)\\
\hline
$\omega\omega$ &  0.02 (1) & 2 (0.5) & 20 ($10^5$) & $1.6 \times 10^3$ (690) \\
\hline
$\phi \phi$ & 0.02 (0.9) & 1.6 (0.3) & 20 ($10^5$) & $1.6 \times 10^3$ (704)\\
 \hline
$\rho \omega$ & 0.5 (23) & 55 (11) & 500 ($2.10^5$) & $4 \times 10^5$ ($1.6 \times 10^5$)\\
\hline
$\rho \phi$ & 0.2 (10) & 22 (4.5) & 230 ($10^5$) & $2 \times 10^5$ ($1.6 \times 10^5$)\\
\hline
$\omega\phi$ & 0.02 (0.9) & 2 (0.4) & 19 ($10^5$) & $1.6 \times 10^3$ (667)\\
\hline
 \hline \hline
\end{tabular}
\end{center}
\caption{\it Double light mesons: The total cross section (number of events/month) for double vector meson production in two-photon reactions in UPC's at RHIC and LHC energies ($t_{min} = 0$ GeV$^2$). Cross sections in units of nb. }
\label{tab2}
\end{table}

Lets now compute the double meson production in ultraperipheral heavy ion collisions within  the BFKL approach. Using Eq. (\ref{dsdtgq}) into Eq. (\ref{sigtot}) and substituting it in Eq. (\ref{sigfoton}) we can calculate the total cross section for the double vector meson production at different values of the center of mass energy and distinct ions. Moreover, using these results  we can estimate the expected number of events for
the RHIC and  LHC luminosities considering the collision of different ions at distinct center of mass energies. In Table \ref{lum}
we present the luminosities and beam energies for $AA$ collisions at RHIC and LHC which we use in our calculations in order to estimate the number of events. Moreover, we assume the  standard $10^6\,s/$month heavy ion run at the RHIC and LHC. Notice that the number of events by year are different for RHIC and LHC, since RHIC is a dedicated heavy ion mode with a run of $10^7\,s$/year and LHC is planned to have one month ($10^6\,s$) of run in its ion mode. Moreover, we have that the center of mass energy at LHC is factor 20 larger than the RHIC one.

In the Tables \ref{tab1}, \ref{tab2}, \ref{tab3} and \ref{tab4} we present our predictions for the production of different combinations of two vector mesons and distinct energies and ions. In particular, in the Tables \ref{tab1} and \ref{tab2} we present our results for double light meson production assuming $t_{min} = 1$ GeV$^2$ and $t_{min} = 0$, respectively. The first aspect which should be emphasized is the large difference between the predictions shown in the two tables. As already shown in Ref. \cite{vic_sauter}, the differential cross sections peaks at small values of $t$, which implies a large contribution from the region of small values of momentum transfer. We have that independently of the choice for the minimum momentum transfer the production of light vector mesons is predicted to be huge, mainly at LHC energies and heavy ion collisions, which become the analysis of this process feasible. In comparison with our previous calculations  for double-$\rho$ production, we have that our predictions using the BFKL approach are somewhat smaller or same  order than obtained in Ref. \cite{double_meson} in terms of the pomeron-exchange factorization relations. However, the comparison depends on the particular choice for the photon parton distribution function (pdf) used in the $\rho J/\Psi$ calculation and on the treatment of the double $J/\Psi$ production. Namely, in the pomeron-exchange factorization relations, $\sigma (\gamma \gamma \rightarrow \rho\rho)\propto [\sigma (\gamma \gamma \rightarrow \rho J/\Psi)]^2/\sigma (\gamma \gamma \rightarrow J/\Psi J/\Psi)$. The result presented here (for $t_{min} = 0$) is equal to prediction in Ref. \cite{double_meson}, $\sigma(AA\rightarrow \rho \rho AA)=200$ $\mu$b, obtained choosing the GRS NLO parameterization for the photon pdf and a NLO BFKL treatment for the double $J/\Psi$ cross section.

In Table \ref{tab3} we present our predictions for the production of double light-heavy mesons. We have that the predicted cross sections for $\rho J/\Psi$ production in Pb-Pb collisions at LHC are in agreement with the results obtained in Ref. \cite{double_meson} using a distinct approach. In Ref. \cite{double_meson}, the $\rho J/\Psi$ cross section at photon-level was computed in the double logarithm approximation (DLLA) of pQCD and its  magnitude depends on the particular choice for the photon pdf. The result presented here is similar to those calculations, $\sigma(AA\rightarrow \rho J/\Psi AA) \approx 1\!-\!2$ $\mu$b, when GRS NLO parameterization for the photon is considered. Finally, in Table \ref{tab4} we present our  predictions for the double heavy vector meson production. We have that the cross sections diminish  for heavier vector mesons. In particular, we have that our result for the double $J/\Psi$ production ($A = Pb$ and LHC energy) is similar the one obtained in Ref. \cite{per1} where we have used a phenomenological analysis also based on the BFKL approach. We have still predicted  a reasonable rate of production, 
 allowing future experimental analyzes, even if the
acceptance for the $J/\Psi$ detection being low. On the other hand, the detection of double $\Upsilon$ production will be a hard task. Due to the higher integrated luminosity for light ions, the event rates are enhanced in this case despite the integrated cross section being small. The present estimate for the double $J/\Psi$ production is consistent with our previous phenomenological analysis using the forward BFKL solution in Ref. \cite{per1}, $\sigma(AA\rightarrow 2J/\Psi AA) \approx 60$ nb. There, the main uncertainty to the overall normalization was the value for the diffractive slope $B_{J/\Psi \,J/\Psi}$, which is clearly determined in the present treatment.

In order to evaluate how much the BFKL resummation amplifies the cross sections, we can compare our predictions which those obtained at Born level. We obtain that  $\sigma_{Born} (AA \rightarrow \rho \rho AA)$ = 6 (736) nb ($t_{min} = 1$ GeV$^2$),   $\sigma_{Born} (AA \rightarrow \rho J/\Psi AA)$ = 0.11 (129) nb ($t_{min} = 0$ GeV$^2$) and 
 $\sigma_{Born} (AA \rightarrow J/\Psi J/\Psi AA)$ = 0.001 (11) nb ($t_{min} = 0$ GeV$^2$) at RHIC (LHC) energies. It is important to emphasize that although the $\gamma \gamma \rightarrow V_1 V_2$ is energy independent at the Born level, there is a dependence on energy which comes from the equivalent photon flux.  We can observe that 
  the BFKL effects imply cross sections  which are a factor 2-3 larger than the Born one, while at LHC this factor is $\ge$ 10.

\begin{table}[t]
\begin{center}
\footnotesize
\begin{tabular}{||c||l|l||l|l||}
\hline \hline
{\bf $V_1V_2$} & {\bf RHIC} & & {\bf LHC} &  \\ \hline 
 & {\bf Si} & {\bf Au} & {\bf Ca} & {\bf Pb} \\\hline
$\rho J/\Psi$ & $8 \times 10^{-3}$ (0.34) & 0.25 (0.05) & 15 ($7 \times 10^{3}$) & 1200 (504)\\
\hline
$\omega J/\Psi$ & $6 \times 10^{-4}$ (0.03) &  0.02 ($4 \times 10^{-3}$) &  1.3 (657) & 101 (42)\\
\hline
$\phi J/\Psi$ & $4 \times 10^{-3}$ (0.18) & 0.08 (0.01) & $2 \times 10^{4}$ ($10^7$) &  $8 \times 10^{4}$ ($3 \times 10^4$) \\
 \hline
$\rho \Upsilon$ & $10^{-6}$ ($5 \times 10^{-5}$) &$3 \times 10^{-6}$ ($6 \times 10^{-7}$) & 0.02 (10) & 1.3 (0.5)\\
\hline
$\omega \Upsilon$ & $10^{-7}$ ($4 \times 10^{-6}$) & $2 \times 10^{-7}$ ($5 \times 10^{-8}$) & $10^{-3}$ (0.85) & 0.1 (0.04)\\
\hline
$\phi \Upsilon$ & $2 \times 10^{-7}$ ($8 \times 10^{-6}$) & $4 \times 10^{-7}$ ($10^{-7}$) & $2 \times 10^{-3}$ (1.5) & 0.2 (0.08)\\
\hline
 \hline \hline
\end{tabular}
\end{center}
\caption{\it Double light-heavy mesons: The total cross section (number of events/month) for double vector meson production in two-photon reactions in UPC's at RHIC and LHC energies. Cross sections in units of nb. }
\label{tab3}
\end{table}

\section{Summary}
\label{conc}
\begin{table}[t]
\begin{center}
\begin{tabular} {||c|c|l|l|l||}
\hline
\hline
& {\bf HEAVY ION}   & $J/\Psi\,J/\Psi$ & $J/\Psi \Upsilon $ & $\Upsilon \Upsilon$  \\
\hline
\hline
 {\bf RHIC} & SiSi & $2 \times 10^{-3}$ ($10^{-2}$) & $1.6 \times 10^{-7}$ ($7 \times 10^{-6}$) & $8 \times 10^{-10}$ ($4 \times 10^{-8}$) \\
\hline
 & AuAu & $2 \times 10^{-3}$ ($5 \times 10^{-4}$)  & $2 \times 10^{-7}$ ($3 \times 10^{-8}$)  & $6 \times 10^{-11}$ ($10^{-11}$)  \\
\hline
\hline
 {\bf LHC} & CaCa & 0.74 (387)& $3 \times 10^{-3}$ (1.8) & $8 \times 10^{-5}$ (0.05)  \\
\hline
&  PbPb & 61 (26) & 0.25 (0.1)& $5 \times 10^{-3}$ ($2 \times 10^{-3}$)  \\
\hline
\hline
\end{tabular}
\end{center}
\caption{\it Double heavy mesons: The total cross section (number of events/month) for double vector meson production in two-photon reactions in UPC's at RHIC and LHC energies. Cross sections in units of nb.}
\label{tab4}
\end{table}

Before to present a summary of our results, lets briefly discuss the main backgrounds for the photon-photon processes in UPC's. 
 An important background  is given by  the photonuclear
 interactions,  since these reactions have similar kinematics.  For the processes considered here,
the diffractive vector meson production in
 photon-pomeron interactions \cite{vicber,kleinvec,per4} should contribute
 significantly. In particular, because the cross section for this process is large, as predicted and recently confirmed by the  experimental result obtained by the  STAR collaboration \cite{star_data},  the probability of having 
double (independent)  production of vector mesons  in a  single nucleus-nucleus collision and associated to multiple interactions,  is non-negligible  \cite{kleinvec}. Moreover,  the two-photon cross sections have been estimated as being at least $10^3$ smaller than the corresponding photoproduction cross sections \cite{klein_vogt}, which make
the experimental separation between the two interactions very
hard.  Our calculations  indicate that the inclusion of the QCD
Pomeron effects implies higher cross sections at two-photon level
and, consequently, larger cross sections in ultraperipheral
collisions. Therefore, the inclusion of these effects  implies that,
in general, the contribution of two-photon interactions is
non-negligible. Moreover, in principle, an
 analysis of the impact parameter dependence should allow to separate between the two classes of reactions, since two-photon interactions can occur at
 a significant distance from both nuclei, while a photonuclear
 interaction must occur inside or very  near a nucleus. 
We notice  that the experimental separation between the two classes of
 processes is an important point, which deserves further precise studies.

As a summary in this paper we have estimated the double vector meson production in ultraperipheral heavy ion collisions within the BFKL approach. We have considered the non-forward solution of the BFKL equation and calculated the total cross section and event rates for different combinations of final states, energies and heavy ions. Our results demonstrate that the number of events predicted at future colliders is large, allowing a future experimental analysis of this process. It is important to emphasize that in our calculations we have considered the leading order solution of the BFKL equation as well as leading order impact factors. In principle, the inclusion of the next-to-leading order corrections implies a smaller growth of the cross section with the energy. Similarly, saturation effects can contribute significantly, reducing the total cross section. Consequently, our predictions should be considered as an upper bound.

\section*{Acknowledgments}
Two of us (M.V.T. Machado and W. K. Sauter) thanks the support of the High  Energy Physics Phenomenology Group, GFPAE IF-UFRGS, Brazil. This work was partially financed by the Brazilian funding agencies CNPq and FAPERGS.

\end{document}